\documentclass{PoS}
\usepackage{natbib}

\title{The first scientific experiment using Global e-VLBI observations: a multiwavelength campaign on the gamma-ray Narrow-Line Seyfert 1 PMN J0948+0022}

\ShortTitle{The first global e-VLBI scientific experiment: MWL campaign on J0948+0022}

\author{\speaker{Marcello Giroletti}$^{a}$, Z. Paragi$^{b,c}$, H. Bignall$^{d}$, A. Doi$^{e}$, L. Foschini$^{f}$, K. Gabanyi$^{g,c}$, J. Blanchard$^{h}$, F. Colomer$^{i}$, X. Hong$^{j}$, M. Kadler$^{k,l}$, M. Kino$^{m}$, H. J. van Langevelde$^{b,n}$, H. Nagai$^{m}$, C. Phillips$^{o}$, M. Sekido$^{p}$, A. Szomoru$^{b}$, A. K. Tzioumis$^{o}$\\
\llap{$^a$}INAF Istituto di Radioastronomia,
	via Gobetti 101, 40129 Bologna (Italy) \\
        E-mail: \email{giroletti@ira.inaf.it}\\
\llap{$^b$}Joint Institute for VLBI in Europe, Postbus 2, 7990 AA Dwingeloo, The Netherlands\\
\llap{$^c$}MTA Research Group for Physical Geodesy and Geodynamics, H-1585 Budapest, Hungary\\
\llap{$^d$}Curtin Institute for Radio Astronomy, Curtin University of Technology, Perth WA 6845, Australia\\
\llap{$^e$}Institute of Space and Astronautical Science, JAXA, 3-1-1 Yoshinodai, Sagamihara, Kanagawa 229-8510, Japan\\
\llap{$^f$}INAF Osservatorio Astronomico di Brera, I-23807 Merate, Italy\\
\llap{$^g$}FOMI Satellite Geodetic Observatory, Budapest, P.O. Box 585, 1592 Hungary\\
\llap{$^h$}Department of Physics, University of Tasmania, Hobart Tasmania 7001, Australia\\
\llap{$^i$}Observatorio Astron\'omico Nacional, E-28014 Madrid, Spain\\
\llap{$^j$}Shanghai Astronomical Observatory, Shanghai 200030, China\\
\llap{$^k$}Dr. Remeis-Sternwarte Bamberg, Sternwartstrasse 7, D-96049 Bamberg, Germany\\
\llap{$^l$}Erlangen Centre for Astroparticle Physics, D-91058 Erlangen, Germany\\
\llap{$^m$}National Astronomical Observatory of Japan, 2-21-1 Osawa, Mitaka, Tokyo, 181-8588, Japan\\
\llap{$^n$}Leiden Observatory, NL 2300 RA Leiden, Netherlands\\
\llap{$^o$}Australia Telescope National Facility, CSIRO, Epping NSW 1710, Australia\\
\llap{$^p$}National Institute of Information and Communications Technology, Kashima Space Research Center, 893-1, Hirai, Kashima, Ibaraki, 314, Japan\\
}

\abstract{The detection of gamma-ray emission by Fermi-LAT from the radio loud Narrow Line Seyfert\,1 PMN\,J0948+0022 (Abdo et al. 2009, ApJ 699, 976) triggered a multi-wavelength campaign between March and July 2009. Given its high compactness (Doi et al. 2006, PASJ 58, 829), inverted spectrum, and 0$^\circ$ declination, the source was an ideal target to observe at 22 GHz with a Global VLBI array extending from Europe to East Asia and Australia. In order to deliver prompt results to be analysed in combination with the other instruments participating in the campaign, the observations were carried out with real time VLBI, for the first time on a Global scale. Indeed, the main results have been published just a few months after the campaign (Abdo et al. 2009, ApJ 707, 727). Here we present additional details about the e-VLBI observations.
}

\FullConference{10th European VLBI Network Symposium and EVN Users Meeting: VLBI and the new generation of radio arrays\\
		September 20-24, 2010\\
		Manchester Uk}

\begin{document}

\section{Introduction}

Narrow line Seyfert 1 (NLS1) are Active Galactic Nuclei (AGN) characterized by unusual optical spectra, with H$\beta$ line FWHM $< 2000$ km/s, a line intensity ratio $[$OIII$]/$H$\beta < 3$, and a bump in FeII \citep{Pogge2000}.
A small fraction of NLS1 is known to be radio loud and in these cases the flat radio
spectra and VLBI variability suggest that several of them could host
relativistic jets \citep{Komossa2006,Doi2006}.
A conclusive evidence for the existence of relativistic jets in these AGN has been obtained thanks to the Large Area Telescope (LAT) on board {\it Fermi}, which has revealed bright gamma-ray emission from the radio loud NLS1 PMN J0948+0022 just a few week after the launch, with flux at $E>200$ MeV is $4.0\pm0.3 \times 10^{-8}$ ph cm$^{-2}$ s$^{-1}$ and photon index $\Gamma = 2.6\pm0.1$ \citep{discovery}, making it one of the brightest gamma-ray AGNs at high latitude \citep{lbas}; using multi wavelength (MWL) data, it has been possible to estimate some basic parameters for this source, such as the size of the emitting region, the magnetic field value, and the particle energy range \citep{discovery,Foschini2010}. A more accurate view could only be attained through a dedicated study based on simultaneous broad band data and therefore a large campaign was organized between March and July 2009, covering gamma-rays, X-rays, UV, optical, and radio; indeed, although radio emission is generally produced somewhat out of the gamma-ray zone, single dish observations can still yield valuable complementary information on correlated flux and spectral variability; in addition, VLBI observations give independent constraints on the jet Doppler factor and viewing angle.

The main results from such extended campaign were reported by \citet{mwl}, while in this contribution we present some additional facts about the global e-VLBI observations specifically organized for this project. Further details and a more extensive interpretation will also be given in a dedicated forthcoming publication.

\section{Observations and Data Reduction}

\begin{figure}
\center 
\includegraphics[width=0.56\textwidth]{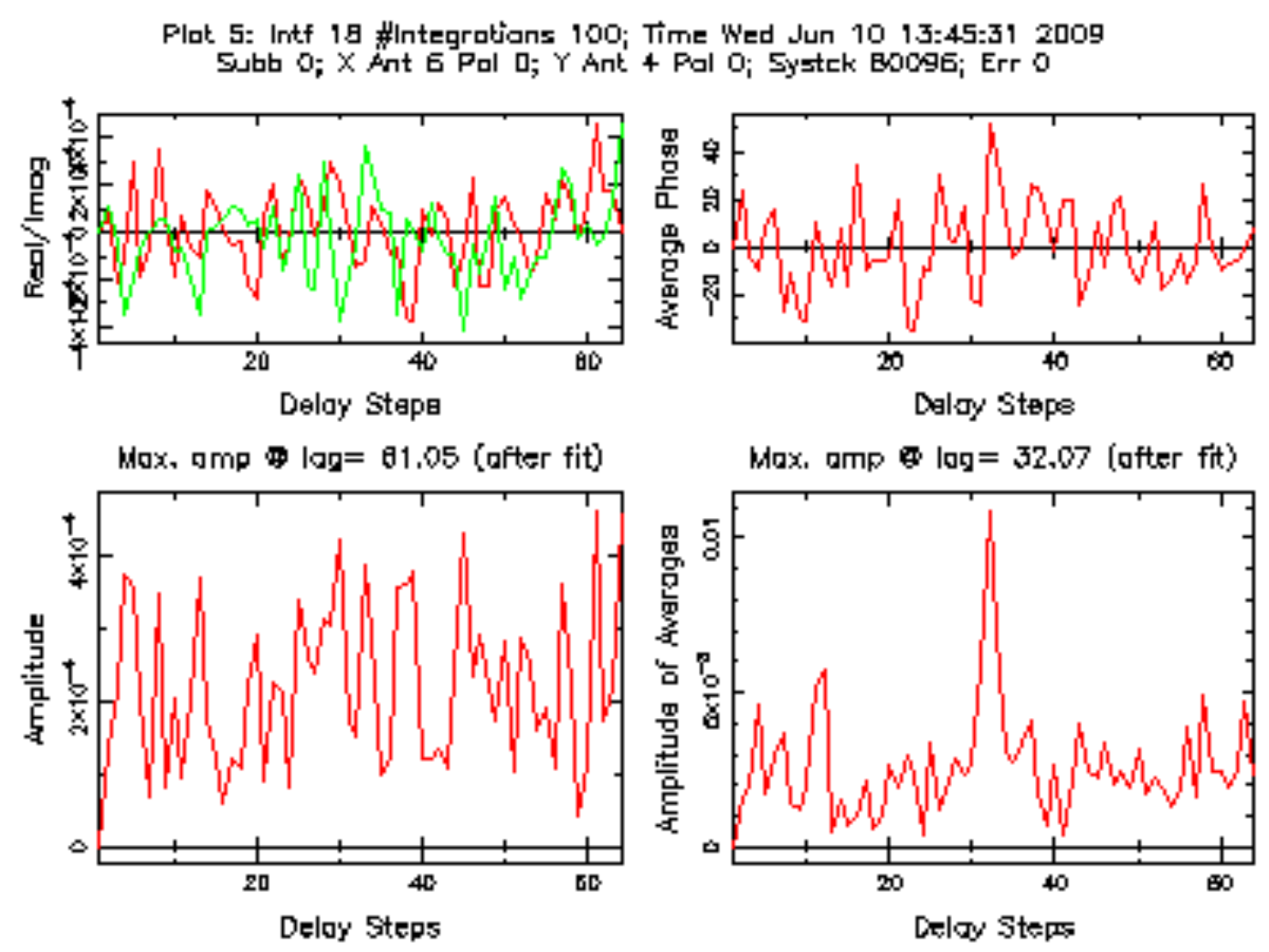}
\includegraphics[width=0.36\textwidth]{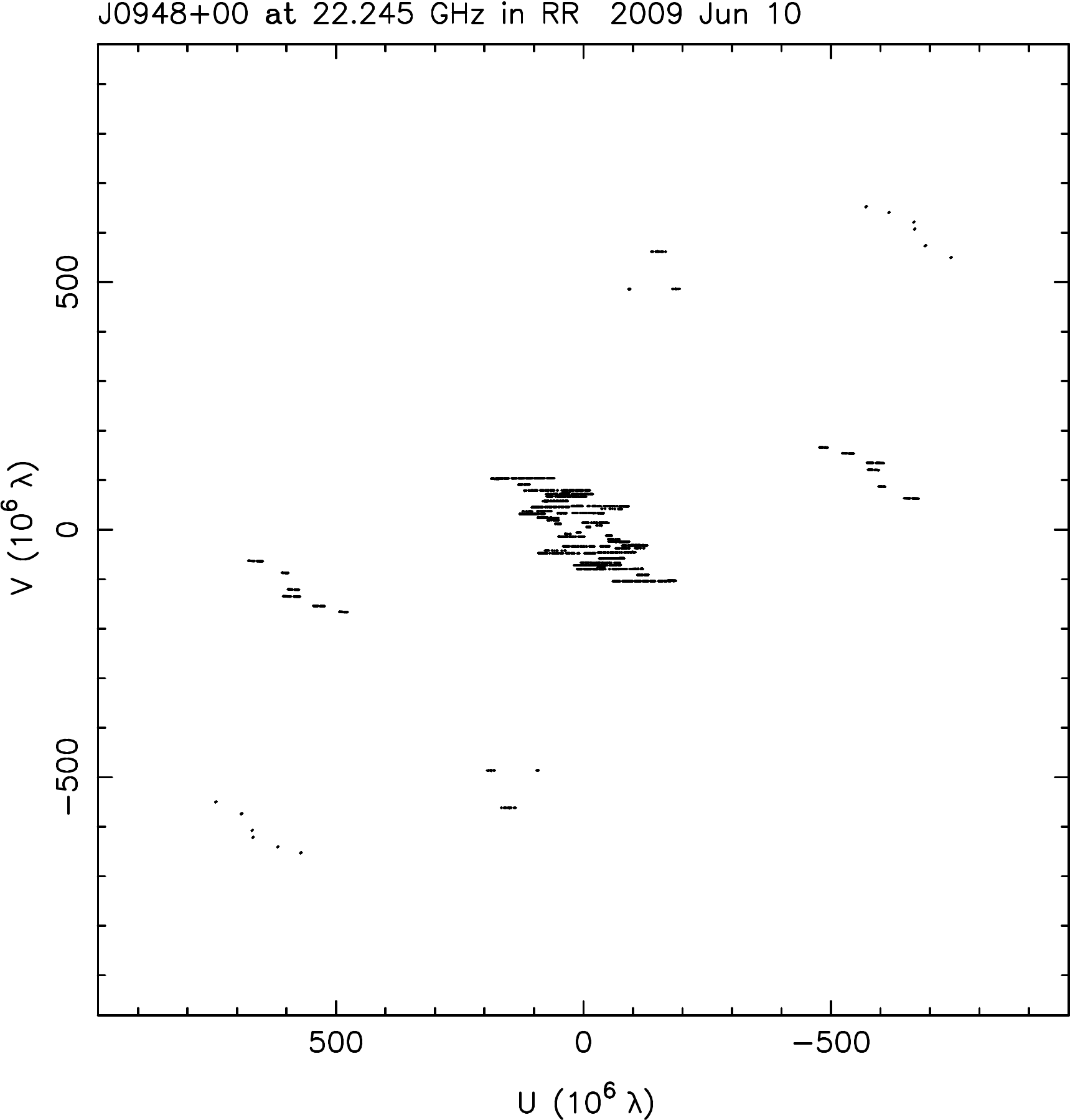}
\caption{The Global e-VLBI experiment EG001B on 2009 June 10. Left: real time fringes between Effelsberg and Medicina; right: coverage of the $(u,v)$-plane.  \label{f.rg001b}}
\end{figure}

A first exploratory short observation was performed for two hours with only European radio telescopes in the 1.6 GHz e-EVN run on 2009 April 21, revealing a 180 mJy point source. On the basis of such detection and of the inverted spectrum revealed by simultaneous single dish observations, 22 GHz e-VLBI observations were deemed doable and scheduled as Target of Opportunity on a global array on 2009 May 23, June 10, and July 4. The participating telescopes were Cambridge, Effelsberg, Jodrell Bank, Medicina, Metsahovi, Onsala, and Yebes in Europe; Kashima and Shanghai in Asia; Hobart, Mopra, Parkes, and the Australia Telescope Compact Array (ATCA) in Australia. A data rate of 512 Mbps was chosen and sustained by most station for the whole duration of the experiment (about 11 hrs). A limited bandwidth was available for a few stations, which however gave good real time fringes. Indeed, despite the different characteristics of the individual parecipating instruments, real time fringes {\bf on the target} have been found to all telescopes in at least one epoch (see Fig.~\ref{f.rg001b}, left panel). Thanks to a mutual visibility of almost one hour between Australia and Europe, baselines exceeding 12000\,km were obtained in the first two epochs, as shown in the $(u,v)$-coverage plot in the right panel of Fig.~\ref{f.rg001b}. In the third epoch, no mutual visibility was obtained due to scheduling constraints of the Australian telescopes.

The visibility data were initially calibrated through the JIVE pipeline, which showed the source was detected, in agreement with the real time results. However, the amplitude calibration information (system temperatures and gain curves) were provided in different times and formats by the individual stations, so a step-by-step data reduction had to be carried out interactively using AIPS and Difmap for producing higher quality results.

\section{Results}

At all epochs, the source was clearly detected and imaged. The angular resolution was around $0.2 \times 0.5$ mas in the first and second epoch and the rms noise level about 1 mJy\,beam$^{-1}$. The image shows a point-like source, although some structure seems to be present in the visibility amplitudes \citep[also in agreement with other published images of the sources, e.g.][]{mwl,Foschini2011}. The total flux density is varying between 300 and 700 mJy; although the exact magnitude of the variability can not be asserted due to a significant uncertainty on the absolute flux scale, it is clear that the source was variable from epoch to epoch. From the angular resolution and the flux density, we obtain a corresponding value for the brightness temperature of $T_\mathrm{B}>3.1\times10^{10}\,$K.

\begin{figure}
\center 
\includegraphics[width=0.5\textwidth]{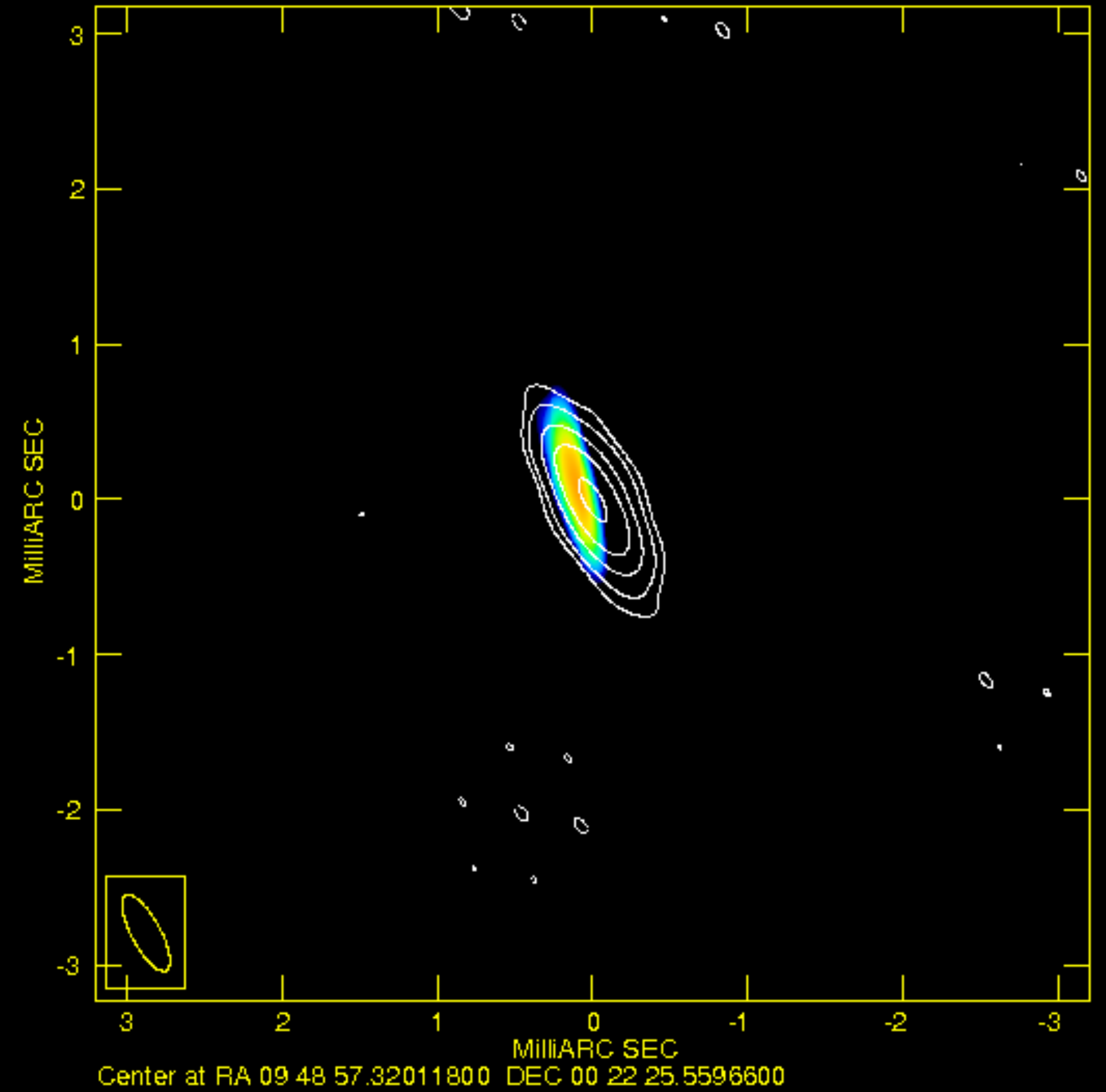}
\caption{Global e-VLBI image of PMN J0948+0022 on 2009 May 23. Contours show total intensity (starting at 4\,mJy\,beam$^{-1}$) and are overlaid to polarized flux in colour scale.  \label{f.source}}
\end{figure}

As the observations were carried out in dual polarization, we could also produce stokes $Q$ and $U$ maps. By combining them, we obtained an image of the source in polarized intensity. Fig.~\ref{f.source} shows it in color scale overlaid to contours of total intensity. Although this image is still preliminary as far as the details are concerned, it is clear that the source is linearly polarized to a few percent level.

\section{Discussion and conclusions}

The census of extragalactic gamma-ray sources has been dominated by blazars (flat spectrum radio quasars and BL Lac type objects) through all the EGRET era. A handful of radio galaxies were also detected. The discovery of gamma-ray emission from radio-loud narrow-line Seyfert1 nuclei is therefore of great importance for the study of relativistic jets in AGNs \citep{Abdo2009d}.

First, it has permitted to seal the issue whether such relativistic structures do exist in radio loud NLS1, as proposed by \citet{Zhou2003} and \citet{Doi2006}. Moreover, thanks to the sensitivity and the surveying capability of {\it Fermi}, it has triggered the organization of a campaign to measure simultaneously in different bands to construct the spectral energy distribution of the source, as well as at studying its variability across the electromagnetic spectrum. In this framework, high frequency Global VLBI observations are of great value, as they permit us to observe the most compact regions of the radio jet above the self-absorption frequency.
Clearly, the maximum return can be obtained with real time VLBI, both because it delivers its results in a suitable time for coordinated multi-wavelength studies and because it permits to observers to monitor the array in real time and directly address problems in the system -- which are likely to arise when a Global array is used.

The overall success of the MWL campaign \citep{mwl} and in particular of the Global e-VLBI observations presented in this contribution is highly encouraging for the continuation of the synergy between high energy astrophysics and VLBI. More generally, Global e-VLBI has proven to be an excellent and reliable tool and we expect it to become a widely used resource in the coming years.

\section*{Acknowledgments}
We acknowledge financial contribution from ASI-INAF I/088/06/0. The European VLBI Network is a joint facility of European, Chinese, South African and other radio astronomy institutes funded by their national research councils.


\begin{thebibliography}{99}

\bibitem[Abdo et al.(2009a)]{discovery} Abdo, A.~A., Ackermann, M., Ajello, 
M., et al.\ \emph{Fermi/Large Area Telescope Discovery of Gamma-Ray 
Emission from a Relativistic Jet in the Narrow-Line Quasar PMN J0948+0022}, 
(2009a) \emph{ApJ}, 699, 976,  [arXiv:0905.4558] 

\bibitem[Abdo et al.(2009b)]{lbas} Abdo, A.~A., Ackermann, M., Ajello, 
M., et al.\ \emph{Bright Active Galactic Nuclei Source List from the First 
Three Months of the Fermi Large Area Telescope All-Sky Survey}, (2009b) 
\emph{ApJ}, 700, 597,  [arXiv:0902.1559] 

\bibitem[Abdo et al.(2009c)]{mwl} Abdo, A.~A., Ackermann, M., Ajello, 
M., et al.\ \emph{Multiwavelength Monitoring of the Enigmatic Narrow-Line 
Seyfert 1 PMN J0948+0022 in 2009 March-July}, (2009c) \emph{ApJ}, 707, 727,  
[arXiv:0910.4540] 

\bibitem[Abdo et al.(2009d)]{Abdo2009d} Abdo, A.~A., Ackermann, M., Ajello, M., et al.\ {\it 
Radio-Loud Narrow-Line Seyfert 1 as a New Class of Gamma-Ray Active 
Galactic Nuclei} 2009, ApJ, 707, L142,  [arXiv:0911.3485] 

\bibitem[Doi et al.(2006)]{Doi2006} Doi, A., Nagai, H., Asada, K., Kameno, 
S., Wajima, K., 
\& Inoue, M.\ \emph{VLBI Observations of the Most Radio-Loud, Narrow-Line Quasar SDSS J094857.3+002225}, (2006) \emph{PASJ}, 58, 829,  [arXiv:astro-ph/0608507] 

\bibitem[Foschini et al.(2010a)]{Foschini2010} Foschini, L., Fermi/Lat Collaboration, Ghisellini, 
G., Maraschi, L., Tavecchio, F., Angelakis, E.\ {\it Fermi/LAT Discovery of 
Gamma-Ray Emission from a Relativistic Jet in the Narrow-Line Seyfert 1 
Quasar PMN J0948+0022} 2010a, ASPC, 427, 243,  [arXiv:0908.3313] 

\bibitem[Foschini et al.(2010b)]{Foschini2011} Foschini, L., Ghisellini, G., 
Kovalev, Y.~Y., et al.\ \emph{The first gamma-ray outburst of a Narrow-Line 
Seyfert 1 Galaxy: the case of PMN J0948+0022 in July 2010}, (2010b) 
\emph{MNRAS submitted}, arXiv:1010.4434

\bibitem[Komossa et al.(2006)]{Komossa2006} Komossa, S., Voges, W., Xu, D., 
Mathur, S., Adorf, H.-M., Lemson, G., Duschl, W.~J., 
\& Grupe, D.\ \emph{Radio-loud Narrow-Line Type 1 Quasars}, (2006) \emph{AJ}, 132, 531,  [arXiv:astro-ph/0603680] 

\bibitem[Pogge(2000)]{Pogge2000} Pogge, R.~W.\ \emph{Narrow-line Seyfert 
1s: 15 years later}, (2000) \emph{NewAR}, 44, 381,  
[arXiv:astro-ph/0005125] 

\bibitem[Zhou et al.(2003)]{Zhou2003} Zhou, H.-Y., Wang, T.-G., Dong, 
X.-B., Zhou, Y.-Y., 
\& Li, C.\ \emph{SDSS J094857.3+002225: A Very Radio Loud, Narrow-Line Quasar with Relativistic Jets?}, (2003) \emph{ApJ}, 584, 147



\end{thebibliography}
\end{document}